\input harvmac

\def\and{ {\rm and} }

\def\be{\begin{equation}}
\def\ba{\begin{eqnarray}}
\def\ea{\end{eqnarray}}   

\def\ee{\end{equation}}

%-------------------
% title page
%-------------------
%
\Title{\vbox{\baselineskip12pt
\hbox{WATPHYS TH-97/03}
%\hbox{gr-qc/96?????}
}}
{\vbox{\centerline{Black Holes of Negative Mass} }}
\baselineskip=12pt
\centerline{
Robert Mann\footnote{$^2$}{Internet:
mann@avatar.uwaterloo.ca}}
\medskip
\centerline{\sl Department of Physics}
\centerline{\sl University of Waterloo}
\centerline{\sl Waterloo, Ontario N2L 3G1 Canada}
\bigskip

\bigskip
\centerline{\bf Abstract}
\medskip
I demonstrate that, under certain circumstances, regions of negative energy
density can undergo gravitational collapse into a black hole. The resultant
exterior black hole spacetimes necessarily have negative mass and non-trivial 
topology.  A full theory of  quantum gravity, in which topology-changing processes 
take place, could give rise to such spacetimes.

%\draftmode
\Date{}

\gdef \jnl#1, #2, #3, 1#4#5#6{ {\it #1~}{\bf #2} (1#4#5#6) #3}

\lref\hawkell{S.W. Hawking and G.F.R. Ellis, {\sl The Large Scale Structure of Spacetime}
(Cambridge University Press, Cambridge, 1973).}
\lref\Bronn{K.A. Bronnikov and M.A. Kovalchuk, J. Phys. {\bf A13} (1980) 187.}
\lref\physblack{J. Kormendy et al., Astrophys. J., Lett. {\bf 473} (1996)  L91. and R. Bender, J Kormendy
and W. Dehnen, Astrophys. J., Lett. {\bf 464} (1996) L123.}
\lref\thorpe{J. Thorpe, {\sl Elementary Topics in Differential Geometry}, (Springer, Berlin,
1979).}
\lref\jolien{J.D. Brown, J. Creighton and R.B. Mann, Phys. Rev. {\bf D50} (1996) 6394.}
\lref\bana{M. Ba\~{n}ados, C. Teitelboim and J. Zanelli, Phys. Rev. Lett. {\bf 69}, 
(1992) 1849.}
\lref\adscmet{R.B. Mann, Class. Quant. Grav. (to be published).}
\lref\israel{W. Israel, Nuovo Cimento {\bf B44} (1966) 1.}
\lref\chase{J.E. Chase, Nuovo Cimento {\bf B67} (1970) 136.}
\lref\amin{S. Aminneborg, I. Bengtsson, S. Holst and P. Peldan, Class. Quant. Grav. 
{\bf 13} (1996) 2707.}
\lref\bala{N. L. Balasz and A. Voros, Phys. Rep. {\bf 143} (1986) 109.}
\lref\wheeler{J.A. Wheeler, Ann. Phys. {\bf 2} (1957) 604.}
\lref\wendy{W.L. Smith and R.B. Mann, University of Waterloo preprint WATPHYS TH-97/02
(gr-qc/9703007).}

The gravitational collapse of a region of sufficiently large density into a black hole
is one of the more striking predictions of general relativity.  It can occur via 
several mechanisms, including stellar collapse, collapse of star clusters, primordial collapse
of a region of sufficiently enhanced density in the early universe, or perhaps
even via quantum fluctuations of the vacuum energy density at sufficiently short 
distances and/or early times.    

A generic feature of black hole formation is the positivity of the initial energy density,
which in turn can collapse into a singularity which is cloaked by an event horizon,
leaving behind an exterior black hole spacetime of positive mass.  A simple inspection of
the Kerr-Newman de Sitter class of metrics indicates that the existence of the
event horizon crucially depends upon the positivity the ADM mass -- a reversal of its
sign yields a spacetime with a naked singularity.

The purpose of this essay is to demonstrate that a region of negative energy density
(which might be produced, say, from quantum fluctuations) 
can also undergo gravitational collapse to a black hole. In order for such a process to
take place, the event horizon of the black hole must be a negatively curved
compact surface. This in turn implies that it is not simply connected, and the
spacetime manifold containing the black hole inheirits a similar topology.

I shall model the region of negative energy with a
stress energy tensor for cloud of freely-falling dust that is given by 
$T_{\mu\nu} = -|\rho| u_\mu u_\nu$, 
where the energy density $\rho < 0$ and $u_\nu$ is the associated four velocity, 
defined such that  $g^{\mu\nu} u_\mu u_\nu = -1$. The form of the metric within the 
dust  is
\eqn\int{
ds^2=-dt^2 + a(t)^2\left( {dr^2\over k r^2+c} + r^2 d\omega^2\right)}
where $t$ is the proper time in the dust cloud, $d\omega^2$ is the metric of a 
compact 2-surface, and $k$ and $c$ are
constants to be determined.  Conservation of stress energy 
implies that $\rho (t) (a(t))^3 = \rho_0 a_0^3 
= -|\rho_0| a_0^3$, where $\rho _0 < 0$ and $a_0$ are the initial density and
scale factor respectively.

The metric exterior to the dust cloud is required (a) to have a negative ADM-type mass
parameter and (b) to have (in its maximal extension) all of its singularities cloaked
by event horizons.  Restricting to the case of static spherical symmetry, it is 
straightforward to show that the only such solutions to the Einstein equations are those with
cosmological constant $\Lambda<0$ and which have the form
\eqn\ext{
ds^2=-\left( {|\Lambda| \over 3}R^2 -1 + {2M\over R}\right) dT^2
+{dR^2 \over ({|\Lambda|\over 3}R^2 -1 + {2M\over R})} +R^2 d\Omega^2}
where $M>0$ and $d\Omega^2 = d\theta^2 + \sinh^2(\theta) d\phi^2$ is the metric of a
2-surface of constant negative curvature.  

Provided $9M^2|\Lambda| < 1$ an event horizon 
will occur for $R=R_+>0$ where
\eqn\outhor{
R_+ = \sqrt{3 \over |\Lambda|} \cos \left({\sin^{-1}(\sqrt{9M^2 |\Lambda|}) \over 3}\right)
-\sqrt{1 \over |\Lambda|} \sin \left({\sin^{-1}(\sqrt{9M^2 |\Lambda|}) \over 3}
 \right) .}
Since the event horizon is a 2-surface (with metric $R_+^2 d\Omega^2$) of
constant negative curvature, it must have a non-zero genus in order for it
to be compact \thorpe . This may be obtained by considering only a portion of the 
hyperbolic surface in the $(\theta , \phi)$ sector, making this section periodic through a
suitable identification of points.  The shape for the section chosen
must be a polygon centered at $\theta=0$ formed from geodesics in the 
$(\theta,\phi)$ sector whose opposite sides are identified.
In order to avoid conical singularities, the angles must sum to more than
$2 \pi$ and the number of sides must be a multiple of four \bala\ .
Since the geodesics on this hyperbolic surface meet at angles smaller than those 
for geodesics meeting on a flat plane, an octagon is the simplest solution, 
yielding a surface of genus 2. In general, a polygon of $4g$ sides
yields a surface of genus $g$, where $g\geq 2$; hence the event horizon 
of the black hole will be a surface of genus $g\geq 2$.

Simple continuity requirements imply that the spacetime manifold with metric \ext\ 
inheirits this identification, and so has topology  $R^2\times H_g$, where $H_g$
is a 2-surface of genus $g\geq 2$.  With this in place,
and using methods for computing quasilocal mass parameters in non-asymptotically 
flat spacetimes \jolien ,  it is straightforward to show from \ext\ that the mass 
of the black hole is $m_{ADM}=-M<0$.  One can therefore regard the exterior spacetime
\ext\ as that corresponding to a black hole of negative mass.  

These negative mass black holes have an inner event horizon at $R=R_-<R_+$, where
\eqn\outhor{
R_- = {2 \over \sqrt{|\Lambda|}} \sin \left({\sin^{-1}(\sqrt{9M^2 |\Lambda|}) \over 3}
 \right).}
The Penrose diagram is similar to that of Reissner-Nordstrom anti de Sitter spacetime.
There is a timelike curvature singularity within the inner horizon at $R=0$. 
If $9M^2|\Lambda| = 1$, then the black hole is extremal, and for $9M^2|\Lambda| > 1$
the spacetime has a naked singularity.

Consider next the process of gravitational collapse of the dust cloud. 
(Collapse of an analogous distribution of positive energy dust into a black hole
of positive mass has been considered in \wendy\ ; the zero-mass case has been
discussed in \amin). Matching the exterior metric \ext\ to the dust metric \int 
using  the matching conditions \israel \chase
\eqn\match{
[g_{ij}]=0   \qquad {\rm and} \qquad   [K_{ij}]=0 }
ensures that the dust edge is a boundary surface with no shell of stress-energy present.
Here $[\Psi]$ denotes the discontinuity of $\Psi$ across the edge, 
$K_{ij}$ is its extrinsic curvature  and the subscripts $i,j$ refer to the coordinates on 
the dust edge.  

The conditions \match{} imply that $d\omega^2=d\Omega^2$, which in turn implies from
the Einstein equations applied within the interior of the dust cloud that $c=-1$,
$k>0$, and
\eqn\dusteq{
\dot a^2 = {-|\Lambda| \over\ 3} a^2 - {8 \over 3} \pi G {a^3_0 |\rho _0| \over a} + k}
where the overdot denotes a derivative with respect to $t$. The constant $k$ is determined
by specifying the initial values of $a$ and $\dot{a}$. In the exterior coordinates 
\ext{} the dust edge is at $R= \Re(t)=r_0 a(t)$, where 
$ds^2 = -dt^2 + \Re^2(t)d\Omega^2$ is the
metric on the edge. A straightforward computation of the matching conditions \match{}
yields equation \dusteq{} with $a(t)\rightarrow \Re(t)/r_0$ -- full consistency then
demands that $M= {4\over 3} \pi G |\rho_0| \Re_0^3$
where $\Re_0=r_0 a_0$ is the initial location of the dust edge.

It would appear that arbitrarily large initial concentrations of
negative energy could collapse to form black holes of arbitarily large negative 
mass $-M$, but this is not the case.  A large enough concentration of negative energy 
will expand outward, so for the dust to collapse from rest to a black hole, 
its initial density must be sufficiently small so that its initial acceleration 
(computed by  differentiating \dusteq{} at $t=0$) is inward.  This implies that 
${4\pi G\rho_0\over \Lambda} < 1$ or alternatively 
$\sqrt{|\Lambda|}\Re_0>(3 M\sqrt{|\Lambda|})^{1/3}$. 
Furthermore, the dust will not collapse to zero size but will instead undergo a bounce
due to its own gravitational repulsion. An analysis of the turning points of the motion
described by \dusteq{} indicates that for a dust cloud collapsing from rest at
$\Re_0>R_+$, the bounce always occurs within the inner horizon $R_-$. Since
$\sqrt{|\Lambda|}R_+ > (3M\sqrt{|\Lambda|})^{1/3}$, a black hole will necessarily form 
provided the magnitude of the initial density is sufficiently small.  This analysis does not 
qualitatively change if the dust cloud has an initial inward velocity.

The evolution of the negative energy dust cloud will be as follows. Starting from
rest, after a finite amount of its own proper time its outer edge will pass through 
$R_+$; this will take an infinite amount of coordinate time as measured by observers
in the exterior spacetime \ext\ .  After a further amount of finite proper time,
it passes through the second horizon and reaches a state of minimum size, and a
timelike singularity forms in the exterior spacetime.  The dust cloud can then
either expand into another universe of similiar topology or possibly form a wormhole 
connected to the same universe. The history of the dust cloud and its formation of 
a black hole is reminiscent of the collapse of a charged dust cloud in which the 
charge is large enough to prevent the occurrence of a singularity inside \hawkell .  
Throughout the process the exterior spacetime is guaranteed to be that given by \ext\  
due to a generalization of Birkhoff's theorem \Bronn .

Under the constraints given on $M$ above,
the collapse of the negative energy cloud leaves behind an exterior black hole spacetime 
whose properties bear some resemblance to those of charged, positive mass anti de Sitter
black holes.  It is straightforward to show that a test body moving toward the black
hole will inevitably fall through the event horizon unless it has sufficiently large
angular momentum.  A consideration of the thermodynamics of the black hole \jolien\
indicates that (apart from the usual redshift factor) its temperature is a decreasing
function of $M=-m_{ADM}$. As the black hole radiates, it will lose energy and $M$
will increase toward its maximal value, the black hole asymptotically approaching 
an extremal state with vanishing temperature.

Where might such black holes enter the panoply of physics? The most likely place
is via topology-changing processes in quantum gravity. It is generally expected
that whatever the final theory of quantum gravity might be, quantum fluctuations in 
geometry should enchance quantum fluctuations in topology \wheeler .  
There is no {\it a-priori} guarantee that such fluctuations always have positive energy;
indeed phenomena such as the Casimir effect indicate quite the contrary.  The exterior
spacetime manifold \ext\ has finite Einstein-Hilbert action \adscmet\ , and so will contribute
in a sum-over-histories approach to quantum gravity.  A more complete understanding
of negative mass black holes and their relevance to physics remain interesting open
questions.

\medskip
\noindent{\bf Acknowledgements}
\smallskip\noindent
The work was supported by the Natural Sciences \& Engineering Research
Council of Canada. 

\listrefs

\end